# YAPAY SİNİR AĞI YÖNTEMİYLE DEPREM TAHMİNİ: TÜRKİYE BATI ANADOLU FAY HATTI UYGULAMASI[1]

Handan ÇAM[2]
Osman DUMAN[3]

**ÖZ**

Gerçekleşecek depremleri önceden kesin bilen, genelleştirilebilecek bir yöntem günümüze kadar geliştirilememiştir. Fakat birçok yöntemle deprem tahmini yapılmaya çalışılmaktadır. Bu yöntemlerden birisi olan Yapay Sinir Ağları, belirlenen girişler ve çıkışlar arasındaki ilişkiyi öğrenerek farklı örüntülere karşı uygun çıkışlar vermektedir. Yapılan bu çalışmada Gutenberg-Richter ilişkisine bağlı ve deprem tahminlerinde kullanılan b değerini temel alan bir ileri beslemeli geri yayılımlı yapay sinir ağı geliştirilmiştir. Türkiye'nin batısında yoğun sismik aktiviteye sahip dört faklı bölgeye ait deprem verileri kullanılarak yapay sinir ağı eğitilmiştir. Eğitim aşamasından sonra aynı bölgeler için daha sonraki tarihlere ait deprem verileri test için kullanılmış ve ağın başarımı ortaya konmuştur. Çalışmada geliştirilen ağın tahmin sonuçları incelendiğinde; ağın gerçekleşmeyecek dediği deprem tahmin sonuçları tüm bölgelerde oldukça yüksek çıkmıştır. Bunun yanında ağın gerçekleşecek dediği deprem tahmin sonuçları, çalışılan bölgeler için belli bir oranda farklı sonuçlar vermiştir.

**Anahtar Kelimeler:** Deprem Tahmini, Yapay Sinir Ağları, İleri Beslemeli Geri Yayılımlı Sinir Ağları.

# EARTHQUAKE PREDICTION WITH ARTIFICIAL NEURAL NETWORK METHOD: THE APPLICATION OF WEST ANATOLIAN FAULT IN TURKEY

**ABSTRACT**

A method that exactly knows the earthquakes beforehand and can generalize them cannot still been developed. However, earthquakes are tried to be predicted through numerous methods. One of these methods, artificial neural networks give appropriate outputs to different patterns by learning the relationship between the determined inputs and outputs. In this study, a feedforward back propagation artificial neural network that is connected to Gutenberg-Richter relationship and that bases on b value used in earthquake predictions was developed. The artificial neural network was trained employing earthquake data belonging to four different regions which have intensive seismic activity in the west of Turkey. After the training process, the earthquake data belonging to later dates of the same regions were used for testing and the performance of the network was put forward. When the prediction results of the developed network are examined, the prediction results that the network predicts that an earthquake is not going to occur are quite high in all regions. Furthermore, the earthquake prediction results that the network predicts that an earthquake is going to occur are different to some extent for the studied regions.

---

**GİRİŞ**

Depremlerin önceden tahmin edilebileceği fikri uzun yıllardır tartışılmaktadır. Deprem tahmini için farklı parametreler oluşturulmaya çalışılmıştır. Bu parametreler depremlerden önce gerçekleşen ve depremi haber veren olaylardır. Anormal hayvan davranışları, gökyüzünde meydana gelen değişimler, yer altı sularında meydana gelen değişimler, akarsu ve denizlerde meydana gelen değişimler ve toprakta bulunan radon gazı yoğunluğundaki değişimler gibi parametreler depremi haber veren parametreler olarak gösterilebilir. Deprem tehlikesinin tahmin edilebilmesi için çeşitli istatistik tabanlı modeller geliştirilmiştir. Bu modeller Poison modeli, Markov modeli, uç değerler dağılımı modelidir. Deprem tehlikesi tahmin modellerinin bazıları büyük depremler için iyi bir tahmin sonucu verirken bazıları küçük ve orta depremlerin tehlikesinin tahmininde iyi sonuçlar vermektedir.

Yapay Sinir Ağları (YSA) tahmin, sınıflandırma gibi problemlerin çözümünde yüksek başarı oranına sahip bir modeldir. Günümüzde bir çok farklı alanda doğrusal ve doğrusal olmayan problemlerin çözümünde başarılı bir şekilde kullanılmaktadır. YSA'ların kendi içerisinde problemin yapısına göre, bağlantı yapılarına göre, öğrenme yöntemine farklı farklı çeşitleri vardır.

YSA modeli ile yapılacak çalışmalarda başarı elde edilebilmesi için girdi parametrelerinin problemi tam temsil edecek şekilde seçilmesi gerekmektedir. Yapılan bu çalışmada depremsellik parametresi olduğu kanıtlanmış olan uzun yıllardır kullanılan Gutenberg Richter yasasına dayanan b parametresi kullanılacaktır. Bu parametre ana parametre olmakla birlikte Omori Utsu kanununa dayanan bir diğer parametrede ağın girişinde kullanılacaktır. Ağın girişinde kullanılacak son parametre ise yine Gutenberg Richter yasasına dayalı 6.0 ve daha büyük şiddette deprem olma olasığılını b parametresine bağlı olarak gösteren parametredir. Ağın çıkışı ise kayıt altına alınan depremden sonraki beş gün içerisinde gerçekleşecek depremin şiddetidir. Ağın ürettiği ve gerçekte gözlemlenen çıktılar daha önce belirlenmiş bir eşik değerin altında ise deprem olmayacağı varsayılmaktadır.

Bu bağlamda çalışmanın amacı; Türkiye'de Gutenberg-Richter ilişkisine bağlı ve deprem tahminlerinde kullanılan b değerini temel alan bir ileri beslemeli geri yayılımlı yapay sinir ağı geliştirerek, ileri tarihli olası depremlerin tahmin edilebilmesidir. Literatürde yer alan yapay sinir ağları ile deprem tahminlemelerinden farklı veriler kullanılacaktır. Diğer bir ifadeyle çalışmada yeni parametreler ve yeni tahminlemeler kullanılacaktır. Bu bakımdan çalışmanın, özellikle Türkiye'de bu konudaki ilk çalışma olacağı ve literatüre yeni bir boyut kazandıracağı düşünülmektedir.

Çalışmada, belirtilen amaç doğrultusunda, 4 farklı bölge üzerinde tahmin işlemi gerçekleştirilmiştir. Birinci bölge olarak Gölhisar Çameli Bölgesi seçilmiştir. Bu bölge Batı Anadolu'da sismik aktivitenin yoğun olduğu bir bölgedir. İkinci bölge olarak Burdur Fay Bölgesi seçilmiştir. Bu bölgeden Batı Anadolu'nun en aktif fayı geçmektedir ve sismik aktivite açısından hareketli bir bölgedir. Üçüncü bölge olarak Büyük ve Küçük Menderes graben bölgesi seçilmiştir. Ege bölgesi ve çevresi düşünüldüğünde tarih boyunca büyük depremler Menderes grabeni üzerinde gerçekleşmiştir. Dördüncü bölge olarak Gediz ve Alaşehir Grabenleri seçilmiştir.

228

Belirlenen amaç ve kapsam çerçevesinde çalışmada, depremlerin tahmini için geçmiş deprem verileri kullanılarak bir YSA modeli geliştirilerek eğitilmiştir. Eğitilen bu modelin ürettiği tahmin sonuçları gerçek sonuçlarla karşılaştırılmış ve ağın başarısı sunulmuştur.

## I. KAVRAMSAL ÇERÇEVE VE LİTERATÜR ÖZETİ
### A. Kavramsal Çerçeve

Yapay sinir ağlarının temeli 1942 yılında McCulloch ve Pitts tarafından ortaya atılan ilk hücre modeliyle başlamıştır. Çalışmalar başlangıçta tıp bilimleri üzerine olmasına rağmen zamanla farklı disiplinlerde de kullanılmaya başlanmıştır(Garip, 2011: 75). Buna göre bir yapay sinir ağı bir çok basit sinir hücresinin birleşiminden meydana gelen kompleks bir sinir ağıdır (Lippman, 1987: 3) ve birbiriyle bağlantılı bir çok sanal nöronun belirli bir yapıda etkileşimiyle oluşur (Zurada, 1992: 15). Yapay Sinir Ağları kendisine verilen veriden öğrenerek kendi kuralları çıkartan (Rojas, 1996: 22) bir yapıya sahiptir. YSA yapay basit sinirlerin birbirlerine farklı etki oranları ile bağlanmasıyla oluşan bir sistemdir. Sinir hücreleri farklı şekillerde birbirleri ile bağ oluşturarak sinir ağı yapısını oluşturur (Haykin, 1992: 16).

1940 öncesi yapılan çalışmalarda mühendislik kullanılmadığından dolayı, YSA'lar üzerinde yapılan mühendislik kökenli ilk çalışmalar; 1940'lı yıllarda McCulloch ve Pitts tarafından yapılan çalışma ile başlamıştır. İlk yapay nöron modeli ortaya atılmış, bu nöronlardan oluşan ağ yapılarının aritmetik ve mantıksal problemlerin çözümünde kullanılabileceği gösterilmiştir ( Mcculloch ve Pitts, 1943:115). 1949 yılında Donald Hebb (1949) tarafından geliştirilen 'hebbian' öğrenme kuralı birçok öğrenme kuralının da temelini oluşturmuştur.

Widrow ve Hoff (1960), basit sinir modelini kullanarak öğrenme gerçekleştirebilen ADALINE (ADAptive LInear NEuron) modelini ortaya atmışlardır. Aynı zamanda ağın eğitimi boyunca toplam hatayı en aza indirmeyi hedefleyen Widrow-Hoff öğrenme kuralını geliştirdiler. MADALINE birden fazla Adaline ünitesinden meydana gelen ağdır 1970'li yıllarda ortaya çıkmıştır (Kaftan,2010).

1969 yılında Minsky ve Pappert yazdıkları 'Algılayıcılar' (perceptrons) adlı kitapta YSA'nın doğrusal olmayan problemlere çözüm üretemediğini ve birçok mantıksal operasyonu (XOR problemi gibi) çözemediğini iddia etmişlerdir. Bu durum 1980'lere kadar YSA çalışmalarında durgunluk yaratmıştır (Öztemel,2003:28). Hopfield (1982) tarafından 1982 yılında doğrusal olmayan ağların geliştirilmesi, YSA'yı duraklama döneminden çıkartmıştır.

### B. Yapay Sinir Ağları ile Deprem Tahmini İle İlgili Yapılan Çalışmalar

Bodri (2001), çalışmasında Magnitüd değeri 6.0'dan büyük depremlerin başlangıç zamanını tahmin etmeye yönelik yapay sinir ağı modeli geliştirilmiştir. İstanbul Teknik Üniversitesi Elektrik-

229

Elektronik ve Maden Fakülteleri 1999 yılında "Elektrostatik Kayaç Gerginlik İzleme Yöntemi ile Deprem Tahmin Sistemi (EKGDT)" isimli bir proje geliştirmeye başlanmışlardır. Projede özel olarak geliştirilmiş tek kutuplu elektrik alan (TEA) ölçüm duyargasından yararlanılarak depremlerin tahmini amacıyla 16 istasyonlu bir ağ kurulmuştur. TEA ölçüm duyargası ile elde edilen verilerin değerlendirilmesinde yapay sinir ağlarından faydalanılmıştır (Özerdem ve Sönmez,2003). Panakkat ve Adeli (2007), matematiksel olarak hesaplanmış sekiz sismik parametrenin analizine dayanarak gelecek aydaki en büyük sismik olayın büyüklüğünü tahmin edecek sinir ağı tasarımı gerçekleştirilmiştir. Tahmin modeli üç farklı sinir ağıyla oluşturulmuş ve başarıları birbiriyle kıyaslanmıştır.

Zhang (2008) deprem tahmin çalışmasında, Geri Yayılımlı Yapay Sinir Ağlarını Genetik Algoritmalar ile birleştirilerek ağın başarım oranının daha yüksek doğruluğa ulaşması sağlamıştır. Wang vd. (2009), çalışmalarında Radyal Tabanlı Fonksiyon(Radial Bias Function) kullanılarak deprem tahmin modeli geliştirmişlerdir. Panakkat ve Adeli (2009) çalışmalarında, büyük bir sismik bölgeyi alt bölgelere bölerek, her bir alt bölge için ve her bir zaman periyodu için depremsellik göstergelerini saymışlardır. Çalışmada, izleyen zaman periyodu süresince o alt bölgede olan en büyük depremin büyüklüğüyle olan ilişkileri Geri Dönüşümlü Sinir Ağı kullanılarak çalışılmıştır. Adeli ve Pannakat (2009), diğer bir çalışmalarında, bir sismik bölgede önceden tanımlanmış gelecek bir zaman dilimi için olacak en büyük depremi tahmin etmek için, sekiz adet sismik gösterge olarak bilinen matematiksel parametre kullanarak olasılıklı bir sinir ağı (PNN) sunmuşlardır. Bu model 4,5 ve 6 arasında büyüklüğe sahip depremler için iyi tahmin doğruluğu vermektedir. Yapılan çalışmada sunulan PNN modeli yazarlar tarafından daha önce geliştirilen ve 6 büyüklüğünden daha büyük depremleri tahmin etmek için iyi sonuçlar vermiş olan Geri Dönüşümlü Sinir Ağı Modelini tamamlamaktadır.

Xu vd. (2010), çalışmalarında, DEMETER uydusu tarafından gözlemlenen veriler kullanılarak deprem oluşumu ve çeşitli faktörler arasındaki bağlantıyı kurmak için yarı uyarlanabilir yapay sinir ağı geliştirmişlerdir. Baltacıoğlu vd. (2010), deprem tahmin çalışmalarının yanında deprem hasarlarının hızlı tespit edilebilmesi amacıyla da yapay sinir ağlarını kullanan modeller geliştirmişlerdir. Geliştirilen yöntemle mevcut binaların hasar durum tespiti yapılmıştır. Alafari vd. (2012), çalışmalarında, Süveyş Körfezi, Akaba Körfezi ve Sina yarım adasını içeren Kuzey Kızıldeniz'de olabilecek depremlerin büyüklüğünü tahmin etmek için kullanılabilecek yapay sinir ağına dayandırılan, yapay zeka tahmin sisteminin uygulanmasını amaçlanmıştır. Diğer yöntemlerle kıyaslandığında yüksek tahmin doğruluğu gösteren yapay sinir yapıları ve farklı yapılandırmalar arasındaki performans değerlendirmesi sunulmuştur. Önerilen şema çoklu gizli katmanlı ileri beslemeli sinir ağı modeline dayalı olarak inşa edilmiştir. Bu model; veri toplama, ön işleme, özellik çıkarma ve sinir ağı eğitimi ve test edilmesinden oluşmuştur. Çalışmada sinir ağı modelinin önerilen diğer yöntemlerden daha yüksek tahmin doğrulu verdiği tespit edilmiştir.

230

Reyes vd. (2013), deprem aktivitesinin yoğun olduğu ülkelerden biri olan Chile'de depremleri tahmin etmek için b değeri, Bath kanunu, Omori Utsu kanunu gibi depremsellikle güçlü ilişkisi olan parametrelere dayanan girdi değerleri kullanılarak belirli bir eşik değerin üstündeki depremleri tahmin eden bir yapay sinir ağı sunmuşlardır. Yapılan tahminler, istatistiksel testler aracılığıyla değerlendirilmiştir ve makine öğrenme sınıflandırıcılarıyla kıyaslanmıştır.

Martínez-Álvarez vd. (2013), çalışmalarında farklı sismik göstergelerin Yapay Sinir Ağları için girdi olarak kullanımını incelemişlerdir. Çalışmada, özellik seçme tekniği uygulaması tarafından farklı sismik bölgelerde başarılı bir şekilde kullanılmış çoklu gösterge kombinasyonu amaçlanmıştır. Orijinal girdi setleri ve farklı sınıflandırıcıların kıyaslanmaları elde edilen başarının derecesini desteklemek için raporlanmıştır. Son olarak farklı sismik göstergelerden elde edilen bilgi, elde etme analizi kullanılarak dört Chile bölgesi ve iki İberia yarımadası bölgesi karakterize edilmiştir.

Kaftan ve Gök (2013), İzmir ve çevresine ait zemin özellikleri yapay sinir ağları kullanılarak incelemişler ve ivmeölçer istasyonlardan gelen veriler ile zemin özellikleri önişleme gerek kalmadan belirlenmiştir. Çelik vd. (2014), Yapay Sinir Ağları ve Destek Vektör Makineleri kullanarak sismik darbelerden deprem tahmini yapmış ve YSA ile %83, Destek Vektör Makineleri ile %91 oranında doğru sınıflandırma bulmuşlardır.

Alexandridis vd. (2014), Radyal Taban Fonksiyonlu Sinir ağlarını kullanarak büyük deprem oluşumlarını tahmin eden bir model geliştirmişlerdir. Yapılan çalışmada California deprem kataloğu kullanılmıştır ve diğer sinir ağı mimarileri ile karşılaştırma yapılmıştır. Zhou ve Zhu (2014), çalışmalarında Levenberg-Marquard geri yayılım sinir ağları kullanılarak deprem tahmin çalışması yapmışlardır.

Inalegwu (2015), Türkiye'de yapay sinir ağları kullanarak ikincil sismik dalgaların geliş zamanını tahmin etmek üzerine bir deprem tahmin araştırması yapmıştır. Çalışmada ikinci dalganın gelişini etkileyen farklı parametreler de ağ modeline katılmıştır. Geliştirilen yapay sinir ağı ile iki sismik dalga arasındaki zaman farkı doğru olarak tahmin edilmiştir.

Iatan (2015), sismik göstergeleri, Olasılıksak Sinir Ağı modelinde girdi parametresi olarak kullanarak deprem tahminine yönelik bir çalışma yapmıştır. Sheng vd. (2015), Çinin Kuzey Sismik Bölgesi'nde Yapay Sinir Ağları kullanılarak deprem tahmini gerçekleştirmişlerdir. Bu tahminler yapılırken büyük bölgeler daha küçük bölgelere bölünmüş ve farklı bölgeler için farklı girdi parametreleri belirlenerek daha iyi sonuçlar alınmıştır.

Bilen vd. (2015), Polonya maden ocaklarından elde edilen verileri KNN (K En Yakın Komşu), SVM, (Destek Vektör Makineleri) ve YSA kullanılarak sınıflandırmışlardır. Çalışmada, depremler %94 başarım oranıyla doğru tahmin edilmiş ve en iyi sonucu KNN metodu vermiştir. Gordan vd. (2016), sinir ağı ve parçacık kolonisi birleşimi aracılığıyla sismik eğim tutarlılığının tahmini üzerine bir model geliştirmişlerdir.

231

**C. YSA Benzeri Yöntemlerle Yapılan Deprem Tahmini ile İlgili Çalışmalar**

Çalışmalarda, YSA'nın analiz yapısına benzer farklı yöntemlerde deprem tahmini yapabilmek amacıyla kullanılmıştır. Bu yöntemlerle yapılan çalışmalar aşağıda özetlenmiştir.

Kulalı (2009) Radon gazı yoğunluğunu, deprem gibi sismik olayların belirlenmesi ve takip edilmesi amacıyla kullanmışlardır. Çalışmada, topraktaki Radon gazı yoğunluğu ölçümü yapılarak ölçüm yapılan bölgedeki depremlerle ilişkilendirilmiştir. Külahçı vd. (2009) Radon gazı ile deprem arasındaki ilişkiyi incelemiştir. Deprem oluşumunda sekiz farklı parametrenin doğrusal olmayan değişimi kullanılmış ve parametreler sunulmuştur. Üç katmanlı ileri beslemeli yapay sinir ağı modelini kullanarak Doğu Anadolu Fay Hattı üzerinde deprem tahmini gerçekleştirmiştir. Yapılan bu çalışma dünya literatürüne girmiştir. Öztürk (2009), Magnitüdü 5 den büyük depremlerin gerçekleştiği bölgelerde artçı şok ve deprem tehlikesi değerlendirilmesi yaptığı çalışmasında, ana şok ile artçı arasındaki ilişkileri hesaplamıştır. Çalışmada deprem öncesi durgun dönem belirlenerek bu dönemin deprem tahmininde kullanımı araştırılmıştır.

Yıldırım (2010), çalışmasında, Türkiye'deki deprem verileri üzerinden veri madenciliği yöntemleri kullanılarak örüntüler çıkartmış ve meydana gelebilecek depremleri önceden tahmin etmeye çalışmıştır. Bu amaçla Türkiye Deprem Portalı geliştirilmiş ve sorgulamalar bu portal üzerinden yapılmıştır. Göker (2010), İzmir Seferihisar bölgesinde topraktaki radon gazı yoğunluğu ölçerek, elde edilen değerleri Normal ve Rayleigh dağılım fonksiyonlarına uygun olarak incelemiştir. İstatistiksel analizler sonucunda Radon gazı yoğunluğu değişimindeki standart sapmalar, meydana gelen depremlerle ilişkilendirilmiştir.

Moustra vd. (2011), sismik elektriksel sinyaller ve zaman serisi Magnitüd verisi kullanılarak deprem tahmin modeli geliştirilmiştir. Ulaş (2011), Çok Düşük Frekanslı (VLF) işaretlerin iyon küre üzerinde iletilirken yaşanan kayıplar üzerinden depremleri tahmin edecek bir algoritma geliştirmiştir. Çalışmada kayıpların oluşmasında neden olan birçok parametre tespit edilmiş ve bu parametreleri içeren veriler üzerinden veri madenciliği yöntemleri kullanılarak anlamlı sonuçlar çıkarılmaya çalışılmıştır. Otari ve Kulkarni (2012), Veri Madenciliği uygulamaları ile deprem tahmini üzerine bir çalışma gerçekleştirmişlerdir. 1989-2011 yılları arasındaki 16 makale analizinin yapıldığı çalışmada deprem tahmini, tsunami tahmini ve sel baskını tahmini için, lojistik modeller, sinir ağları, ve karar ağaçları kullanılmıştır. Bodur (2012), YSA'a alternatif olarak deprem konumlarının belirlenmesinde bulanık mantık yaklaşımını kullanmıştır.

Raoff Nasser (2012), deprem tahmininde depremlerden önce gözlemlenen basınç birikimini kesme dalgası ayrımı analizi test etmiştir. Bu analizinin yapılabilmesi için MATHLAB ortamında kesme dalgası ayrım parametrelerini kullanan bir program geliştirilmiştir. Alafari vd. (2012), 1975 yılında gerçekleşen Haicheng depremindeki geoelektriksel ölçüleri tahmin etmişlerdir. Bu deprem bilim adamları tarafından tahmin edilen ilk yüksek magnitüd değerine sahip depremdir. Ancak bu

232

depremden bir yıl sonra Tangshan depremi bilim adamları tarafından tahmin edilememiştir. Bu depremde yaklaşık 250 000 kişi hayatını kaybetmiştir.

## II. METEDOLOJİ
### A. Yöntem ve Veri Seti

Çalışmada, Türkiye'nin Batı Anadolu fay hattı içindeki 4 farklı bölge üzerinde tahmin işlemi gerçekleştirilecektir. Bu bölgelere ait bilgiler Tablo 1'de gösterilmiştir.

**Tablo 1. Tahmin Yapılan Bölgeler**

| Bölge Adı | Konumu | Bölge Numarası |
|---|---|---|
| Gölhisar Çameli Bölgesi | Batı Analdolu | Bölge 1 |
| Burdur Fay Bölgesi | Batı Anadolu | Bölge 2 |
| Büyük ve Küçük Menderes Graben Bölgesi | Batı Anadolu | Bölge 3 |
| Gediz ve Alaşehir Graben Bölgesi | Batı Anadolu | Bölge 4 |

Türkiye'nin Batı Anadolu'sunda deprem tahmini için Yapay Sinir Ağı konfigürasyonu gerçekleştirilmiştir. Bu doğrultuda Gutenber Richter tarafından önerilen denklemdeki b değeri depremselliğin tanımında parametre olarak kullanılmıştır. b değeri farklı yöntemler kullanılarak hesaplanabilmektedir. b değeri hesabında en sık kullanılan yöntem en büyük olasılık yöntemidir. Çalışmada, YSA ile tahmin için kullanılacak depremsellik parametresi olarak b değeri alınmıştır. Çalışma kapsamında belirlenen bölgeler için b değerleri arasındaki fark yani değişim YSA'nın eğitiminde ve test edilmesinde girdi verisi olarak kullanılmıştır.

Ağı eğitmen ve test etmek için gerekli olan veriler Öztürk (2015)'den elde edilmiştir. İlgili bölgeler için elde edilen veriler, 2013 yılı sonuna kadar gerçekleşmiş deprem kayıtlarıdır. Deprem kayıtlarındaki magnitüd değeri kullanılarak b değeri hesaplanmıştır. Ağda kullanılan ilk beş girdi parametresi b değerleri arasındaki değişimlerdir. Ağın altıncı girdi parametresi artçı depremlerin zamanla azalım oranını gösteren Omori Utsu yasasına dayanır. Bu parametre ana şoktan önceki yedi gün içerisinde meydana en büyük depremin magnitüd değeridir. Ağın girişinde kullanılan son girdi parametre ise Gutenberg Richter yasası temeline dayanan 6.0 ve üstünde deprem olma olasılığını b değerine bağlı olarak gösteren parametredir. Ağın ürettiği çıktı değeri, girdi setinde kullanılan depremden sonraki beş gün içerisinde gerçekleşecek depremin magnitüd değeridir.

Çalışmada girdi sayısı, katman sayısı ve bu katmanlarda ki nöron sayıları belirlendikten sonra ağda eğitim aşamasına geçilmiştir. YSA'da katmanlardaki nöronlar arasında bağlantı ağırlıklarının belirlenmesi işlemi eğitim olarak isimlendirilir. Ağda giriş verileri ve çıkış verileri kullanılarak ağırlıkların problemi temsil edecek ideal değerlere ulaşması sağlanmıştır. Ağın bu şekilde genelleştirme yapacak duruma gelmesi öğrendiğini göstermektedir. Eğitim süreci tamamlandıktan sonra geliştirilen YSA modeli çıktı vermeye hazır hale gelmiş ve ağın daha önce karşılaşmadığı

233

örnekler ağa girdi olarak verilerek ağın çıktı üretmesi sağlanmıştır. Verilerin temizlenmesi ve parametrelerin oluşturulması aşağıda açıklanmıştır.

### B. Verilerin Temizlenmesi

Bölgelere ait sismik veriler 2014 yılı başlangıcına kadar dat uzantılı olarak boylam, enlem, yıl, ay, gün, magnitüd, derinlik, saat, dakika, süre bilgileri aralarında bir boşluk olacak şekilde satırlarda saklanmaktadır. Veriler üzerinde işlem yapmak için tüm veriler Excel programına aktarılarak ilgili alanlar sütunlara çevrilmiştir. Excel programı kullanılarak bölgelere ait 2000 yılı sonrasına ait veriler filtrelenmiştir. Bölgelere ait kesme magnitüd değerlerinden düşük veriler de filtrelenmiş böylece veriler girdi parametreleri çıkartılacak hale getirilmiştir. Ayrıca katalog verilerinde tarihler gün ay yıl farklı sütunlarda yer almaktaydı. Tarih ile ilgili gün ay yıl sütunları birleştirilerek tarihe dönüşüm yapılmıştır. Kullanılan katalog içerisinde eksik bilgi bulunmamaktadır. Tüm parametrelerin değerleri vardır, boş veya eksik veri yoktur.

### C. Parametrelerin Elde Edilmesi

Çalışmada, girdi ve performans parametrelerinin hesaplanmasında Reyes vd. (2013) tarafından önerilen yöntem kullanılmıştır. Çalışma yapılan bölgede bir deprem olduğunda bu deprem bilgisiyle ilgili yedi giriş ve bir çıkışa sahip yeni bir vektör oluşturulmuştur. Girdi vektörü, girdi parametrelerini temsil etmektedir. Çıktı vektörü ise çıktı parametresini temsil etmektedir. Eğitim ve test vektörleri girdi ve çıktı parametrelerinin toplamını içermektedir. Bunların hesaplanmasında Excel programına aktarılan katalog verileri kullanılmıştır.

#### 1. Girdi Parametrelerinin Elde Edilmesi

Katalog verileri Excel programına aktarılıp gerekli düzenlemeler yapıldıktan sonra 7 adet girdi parametresi hesaplanmıştır. İlk beş girdi değerinin hesaplanmasında Gutenberg Richter yasası ile ilişkili b değerleri kullanılmıştır. b değeri hesabında hesaplama yapılacak bölgedeki belirli sayıdaki veya belirli tarih aralığındaki kaydedilmiş deprem verisi kullanılmıştır. Yapılan çalışmada 50 deprem verisinden bir b değeri hesaplanmıştır. Bu denklem aşağıda verilmiştir.

$$b_i = \frac{\log(e)}{\left(\frac{1}{50}\right)\sum_{j=0}^{49} M_{i-j} - Mc} \qquad (1)$$

Denklem 1'deki $M_i$, i. depremin magnitüd değeridir. Mc değeri ise kesme magnitüd değeridir. Çalışma yapılan dört farklı bölgede farklı kesme magnitüdleri kullanılmıştır. Kesme değeri Bölge 1 Gölhisar Çamelinde "3,0", Bölge 2 Burdur fay zonunda "2,8", Bölge 3 Büyük küçük menderes bölgesinde "2,9", Bölge 4 Gediz Alaşehir graben bölgesinde "2,8" olarak alınmıştır.

b değeri hesaplaması yapılmadan önce Excel programına aktarılan ve tüm deprem kayıtlarını içeren veriler, parametresi hesaplanacak bölgenin kesme magnitüd değerine göre filtrelenmiştir ve

234

kesme magnitüd değerinden küçük olan deprem kayıtları silinmiştir. Böylece çok küçük değerli deprem verilerinin deprem tahmin modeline yapacağı olumsuz etkilerden kaçınılmıştır.

b değeri hesaplamasında, birinci b değeri hesaplanırken katalog verisinde 1-50 aralığındaki deprem verilerinin magnitüd değerleri kullanılmıştır. İkinci b değeri hesabında 2-51 ve üçüncü b değeri hesabında 3-52 aralığındaki deprem verilerinin magnitüd değerleri kullanılmıştır. Hesaplanan b değerleri arasındaki fark YSA'da girdi parametresi olarak kullanılacaktır. b değerleri arasındaki değişimler Δb ile gösterilmiştir. Delta değişimi, girdi parametresi olan x değerlerine denk alınmıştır.

$$\Delta b_{1i} = b_i - b_{(i-4)} \equiv x_{1i} \quad (2)$$

$$\Delta b_{2i} = b_{(i-4)} - b_{(i-8)} \equiv x_{2i} \quad (3)$$

$$\Delta b_{3i} = b_{(i-8)} - b_{(i-12)} \equiv x_{3i} \quad (4)$$

$$\Delta b_{4i} = b_{(i-12)} - b_{(i-16)} \equiv x_{4i} \quad (5)$$

$$\Delta b_{5i} = b_{(i-16)} - b_{(i-20)} \equiv x_{5i} \quad (6)$$

Denklemlerden anlaşılacağı gibi x girdi parametresinin ilk değerinin hesaplanabilmesi için asgari 70 adet b değerinin hesaplanması gerekmektedir. 70 deprem verisinin ilk 50 tanesi bir b değeri hesabı için kullanılmıştır. b değerlerindeki değişimlerin tamamının hesaplanabilmesi için en az 20 adet b değeri gerekmektedir. Ağın girdi parametrelerinin ilk beş tanesi denklem 2,3,4,5 ve 6 kullanılarak elde edilmiştir. Altıncı girdi parametresi olan $x_{6i}$ tahmin yapılacak bölge içerisindeki girdi olarak kullanılan depremden önceki 7 gün içerisinde kayda alınan en büyük depremin magnitüd değeridir. Bu parametre dolaylı olarak Omori/Utsu ve Bath kanunlarına dayanan bilgileri YSA'ya sağlamaktadır. Altıncı girdi parametresinin matematiksel gösterimi denklem 7'de gösterilmiştir.

$$x_{6i} = \max\{M_d\}, t \in [-7, 0) \quad 7$$

Yedinci girdi parametresi olan $x_{7i}$ 6.0 ve üzerinde magnitüde sahip deprem olma olasılığını tanımlamaktadır. Bu bilginin girdi olarak eklenmesi Gutenberg-Richter yasasının dinamik bir şekilde kapsanması içindir. Olasılık yoğunluk fonksiyonundan hesaplanmıştır. Yedinci girdi parametresi matematiksel gösterimi denklem 8'de gösterilmiştir.

$$x_{7i} = P(M_d \geq 6.0) = e^{(-3b_i/\log(e))} = [10]^{(-3b_i)} \quad 8$$

235

### 2. Çıktı Parametresinin Elde Edilmesi

Ağın çıktı parametresi bir tanedir ve y_i ile temsil edilir. Çalışma yapılan bölge içerisinde kesme magnitüdü üzerinde değere sahip bir depremden sonraki beş gün içerisinde ölçülen en büyük M_d değeridir. y_i değeri kesme magnitüdünden büyük değere sahiptir. Deprem olmamışsa veya kesme magnitüdünden düşük şiddette deprem olmuşsa 0 alınır. Çıktı parametresinin matematiksel hesaplaması denklem 9'da gösterilmiştir.

$$Y_i = \max\{M\_d\}, t \in (0,5] \qquad 9$$

Çıktı parametresi hesabında hesaplama manuel olarak yapılmıştır.

### D. Yapay Sinir Ağı Özellikleri

Yapılan çalışmada YSA modeli Mathlab programı kullanılarak gerçekleştirilmiştir. Kullanılan YSA'nın özellikleri Tablo 2'de sunulmuştur.

**Tablo 2. Yapay Sinir Ağı Özellikleri**

| Parametreler | Değerler |
| --- | --- |
| Girdi Parametreleri | 7 |
| Gizli Katman Nöron Sayısı | 15 |
| Çıktı Nöron Sayısı | 1 |
| Aktivasyon Fonksiyonu | Sigmoid |
| Ağ Topolojisi | İleri Beslemeli |
| Öğrenme Paradigması | Geri Yayılımlı |

### III. BULGULAR

### A. Çalışma Bölgelerine Ait Magnitüd Analizleri

Çalışma kapsamındaki dört farklı bölgeye ait ve 2000 yılından sonrasına ait katalog verileri üzerinden temel deprem istatistikleri incelenmiştir. Kayıt altına alınan depremlerin bölgelere ait magnitüd değerlerleri göre sayısı aşağıdaki grafiklerde gösterilmiştir.

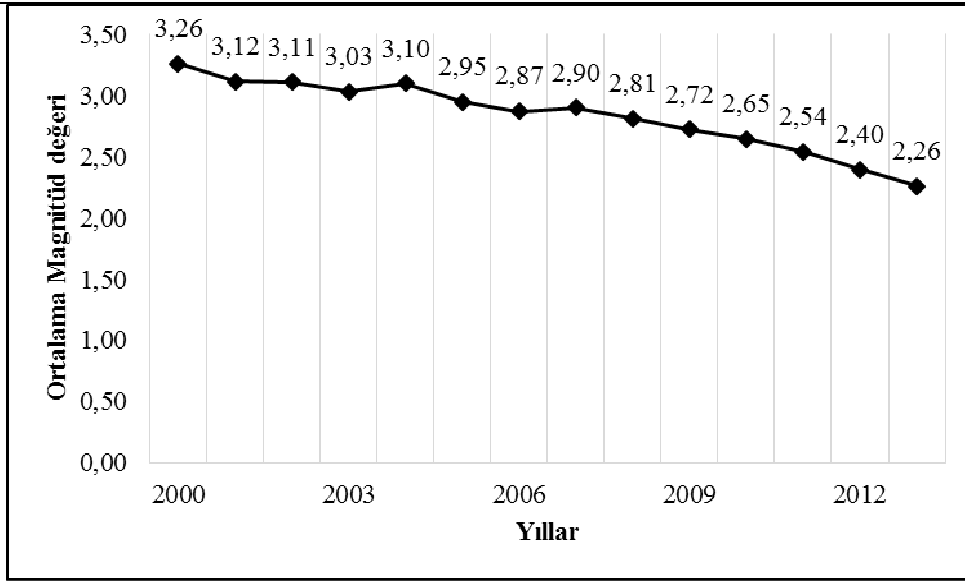

**Grafik 1. Bölge 1 Ortalama Deprem Magnitüd Değerleri**

Grafik 1'de Gölhisar Çameli Bölgesi'nde 2000 yılı sonrasına ilişkin ortalama deprem magnitüdleri gösterilmiştir. Grafik incelendiğinde bu bölgede yıllar geçtikçe ortalama deprem magnitüd değerinin düştüğü görülmektedir.

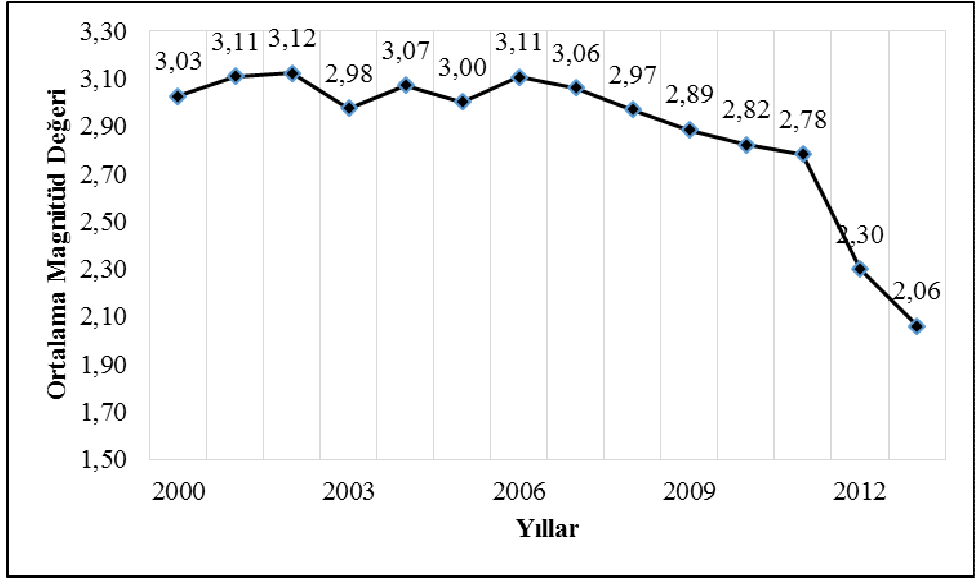

**Grafik 2. Bölge 2 Ortalama Deprem Magnitüd Değerleri**

Grafik 2'de Burdur Fay Bölgesi'nin 2000 yılı sonrasında yıllara göre ortalama deprem magnitüdleri gösterilmiştir. Grafik incelendiğinde çalışma bölgesinde yıllar geçtikçe ortalama deprem magnitüd değerinin düştüğü ve en son 2,06 olarak hesaplandığı görülmektedir.

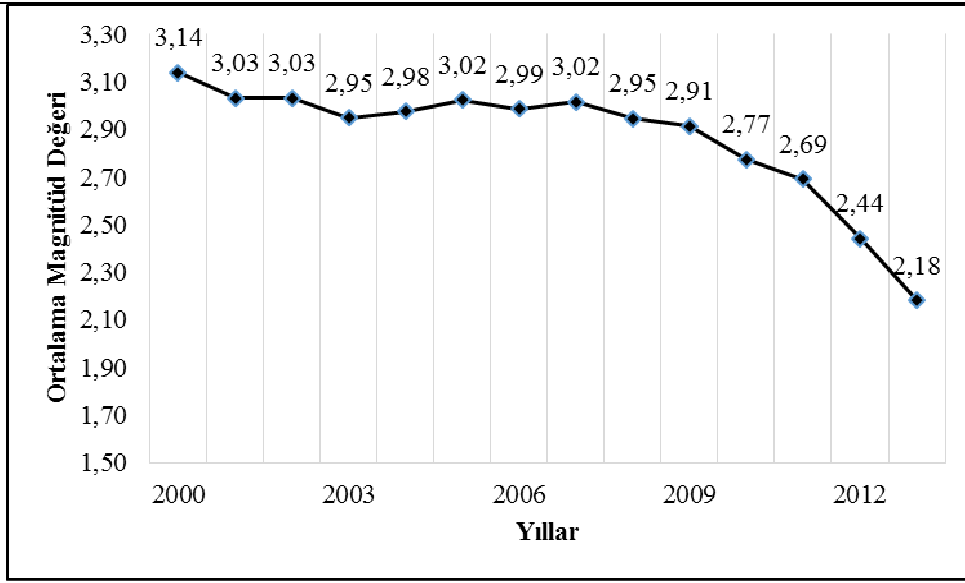

**Grafik 3. Bölge 3 Ortalama Deprem Magnitüd Değerleri**

Grafik 3'de Büyük ve Küçük Menderes Bölgesi'nin 2000 yılı sonrasında hesaplanmış ortalama deprem magnitüdleri gösterilmiştir. Önceki iki bölgede olduğu gibi bu bölgede de yıllar geçtikçe ortalama deprem magnitüd değerinin düştüğü görülmektedir. Bu yıllarda en yüksek magnitüd değeri 3,14, en düşük magnitüd değeri ise 2,18 olarak hesaplanmıştır.

238

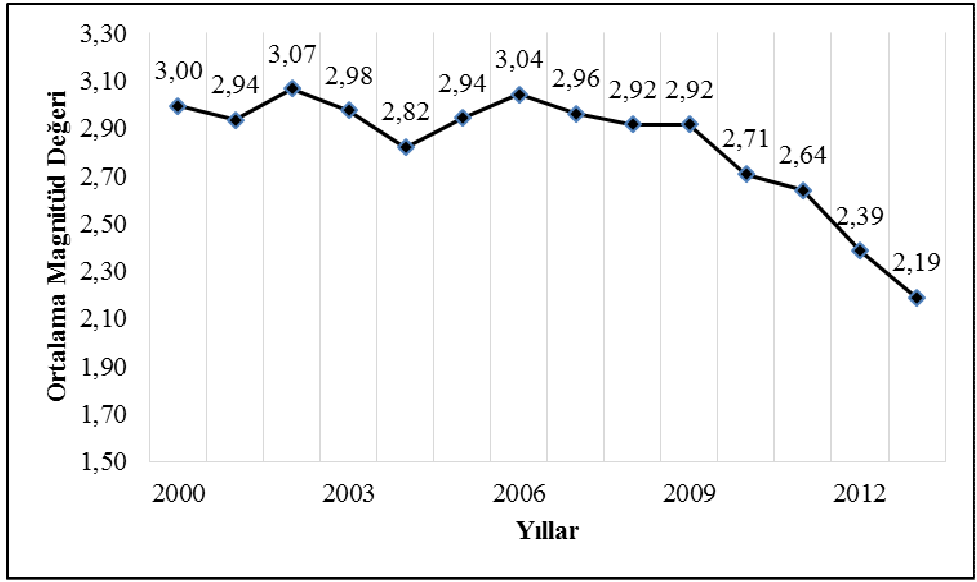

**Grafik 4. Bölge 4 Ortalama Deprem Magnitüd Değerleri**

Gediz ve Alaşehir Graben Bölgesi ait ortalama deprem magnitüd değerleri Grafik 4'de gösterilmiştir. Grafik incelendiğinde çalışma bölgesinde son yıllarda ortalama deprem magnitüd değerinin dalgalı bir seyir izlediği ancak genel itibariyle yıllara göre azaldığı görülmektedir. Bu bölgede en yüksek magnitüd değeri 3,07, en düşük magnitüd değeri ise 2,19 olarak hesaplanmıştır.

**B. YSA Eğitim Testi Sonuçları**

Çalışmada, Gölhisar Çameli Bölgesi için 1 Kasım 2007 ve 25 Ekim 2010 tarihleri arasında magnitüd değeri 3.0 ve üstünde olan 122 adet deprem kaydı YSA'da eğitim amacıyla kullanılmıştır. Eğitim süreci 500 devir (epoch) ile tamamlanmıştır. Ağın eğitimi sonucunda eğitim verileri ağa verildiğinde (self test) ortaya çıkan ve Gölhisar Çameli Bölgesine ait ağın performansı Tablo 3'de gösterilmiştir.

**Tablo 3. Gölhisar Çameli Bölgesi YSA Eğitim Sonuçları**

| Parametre | Değer | Yüzde |
|---|---|---|
| TP | 2 | |
| TN | 101 | |
| FP | 7 | |
| FN | 12 | |
| $P_0$ | 0,8938053 | 89,38 |
| $P_1$ | 0,2222222 | 22,22 |
| $S_n$ | 0,1428571 | 14,29 |
| $S_p$ | 0,9351852 | 93,52 |
| Ortalama | | 54,85 |

Ağın eğitiminde $P_0$ değeri yaklaşık olarak %90 çıkmaktadır. Bu değer istenilen aralıktadır. Eğitim sonucu eğitim verisi ağa verildiğinde ağ; 2 depremin olacağını doğru tahmin etmiştir. 101 depremin gerçekleşmeyeceğini doğru tahmin etmiştir. Buna karşılık 12 depremi tahmin edememiş ve 7 depremi yanlış tahmin etmiştir.

Burdur Fay Bölgesinde 3 Ocak 2006 tarihinden 25 Mart 2009 tarihine kadar magnitüd değeri 2.8 ve üstünde olan 122 adet deprem kaydı, YSA'da eğitim amacıyla kullanılmıştır. Eğitim süreci 500 devir (epoch) ile tamamlanmıştır. Ağın eğitim performansı Tablo 4'de gösterilmiştir.

**Tablo 4. Burdur Fay Zonu YSA Eğitim Sonuçları**

| Parametre | Değer | Yüzde |
|---|---|---|
| TP | 1 | |
| TN | 100 | |
| FP | 3 | |
| FN | 18 | |
| $P_0$ | 0,847458 | 84,75 |
| $P_1$ | 0,25 | 25,00 |
| $S_n$ | 0,052632 | 5,26 |
| $S_p$ | 0,970874 | 97,09 |
| Ortalama | | 53,02 |

239

Ağın eğitiminde P0 değeri yaklaşık olarak %85 çıkmaktadır. Bu değer istenilen aralıktadır. Eğitim sonucu eğitim verisi ağa verildiğinde ağ; 1 depremin olacağını doğru tahmin etmiştir. 100 depremin gerçekleşmeyeceğini doğru tahmin etmiştir. Buna karşılık 18 depremi tahmin edememiş ve 3 depremi yanlış tahmin etmiştir.

Büyük ve Küçük Menderes Bölgesinde eğitim amacıyla 10 Mart 2010 ve 11 Ocak 2011 tarihleri arasında magnitüd değeri 2.9 ve üstünde olan 122 adet deprem kaydı YSA'da kullanılmıştır. Eğitim süreci 500 devir (epoch) ile tamamlanmıştır. Ağın eğitimi sonucunda eğitim verileri ağa verildiğinde ağın performansı Tablo 5'de gösterilmiştir.

**Tablo 5. Büyük ve Küçük Menderes Bölgesi YSA Eğitim Sonuçları**

| Parametre | Değer | Yüzde |
|---|---|---|
| TP | 14 | |
| TN | 79 | |
| FP | 8 | |
| FN | 21 | |
| $P_0$ | 0,79 | 79,00 |
| $P_1$ | 0,636364 | 63,64 |
| $S_n$ | 0,4 | 40,00 |
| $S_p$ | 0,908046 | 90,80 |
| Ortalama | | 68,36 |

Ağın eğitiminde $P_0$ değeri yaklaşık olarak %80 çıkmaktadır. Bu değer istenilen aralıktadır. Eğitim sonucu eğitim verisi ağa verildiğinde ağ 14 depremin olacağını doğru tahmin etmiştir. 79 depremin gerçekleşmeyeceğini doğru tahmin etmiştir. Buna karşılık 21 depremi tahmin edememiş ve 8 depremi yanlış tahmin etmiştir.

Gediz ve Alaşehir Graben Bölgesi eğitim için 3 Aralık 2007 ve 10 Mayıs 2010 tarihleri arasına magnitüd değeri 2.8 ve üstünde olan 122 adet deprem kaydı YSA'da kullanılmıştır. Eğitim süreci 500 devir (epoch) ile tamamlanmıştır. Ağın eğitimi sonucunda eğitim verileri ağa verildiğinde ağın performansı Tablo 6'da gösterilmiştir.

**Tablo 6. Gediz ve Alaşehir Graben Bölgesi YSA Eğitim Sonuçları**

| Parametre | Değer | Yüzde |
|---|---|---|
| TP | 2 | |
| TN | 98 | |
| FP | 2 | |
| FN | 20 | |
| $P_0$ | 0,830508 | 83,05 |
| $P_1$ | 0,5 | 50,00 |
| $S_n$ | 0,090909 | 9,09 |
| $S_p$ | 0,98 | 98,00 |
| Ortalama | | 60,04 |

Ağın eğitiminde $P_0$ değeri yaklaşık olarak %84 çıkmaktadır. Bu değer istenilen aralıktadır. Eğitim sonucu eğitim verisi ağa verildiğinde ağ 2 depremin olacağını doğru tahmin etmiştir. 98 depremin gerçekleşmeyeceğini doğru tahmin etmiştir. Buna karşılık 20 depremi tahmin edememiş ve 2 depremi yanlış tahmin etmiştir.

**C. YSA Uygulama Sonuçları**

Ağın eğitimi tamamlandıktan sonra YSA'nın test işlemleri yapılmıştır. YSA'nın bölgelere göre ayrı ayrı gerçekleştirilen uygulamasında elde edilen parametreler ve tahmin sonuçları aşağıda açıklanmıştır. Gölhisar Çameli Bölgesi 31 Ekim 2010 ile 28 Aralık 2013 tarihleri arasındaki magnitüd değeri 3.0 ve üzerinde olan 122 adet deprem verisi çalışmada YSA'nın test edilmesi amacıyla kullanılmıştır. Ağın test sonucu performans parametreleri Tablo 7'de gösterilmiştir.

**Tablo 7. Gölhisar Çameli Bölgesi YSA Test Sonuçları**

| Parametre | Değer | Yüzde |
|---|---|---|
| TP | 1 | |
| TN | 86 | |
| FP | 5 | |
| FN | 30 | |
| $P_0$ | 0,741379 | 74,14 |
| $P_1$ | 0,166667 | 16,67 |
| $S_n$ | 0,032258 | 3,23 |
| $S_p$ | 0,945055 | 94,51 |
| Ortalama | | 47,13 |

Test sonucunda $P_0$ değeri yaklaşık olarak %74 çıkmaktadır ve bu değer istenilen aralıktadır. Ağ, 1 depremin olacağını ve 86 depremin gerçekleşmeyeceğini doğru tahmin etmiştir. Buna karşılık 30 depremi tahmin edememiş ve 5 depremi yanlış tahmin etmiştir.

Diğer çalışma bölgemiz olan, Burdur Fay Bölgesinin 7 Nisan 2009 ile 19 Aralık 2013 tarihleri arasındaki magnitüd değeri 2.8 ve üzerinde olan 122 adet deprem verisi YSA'nın test edilmesi amacıyla kullanılmıştır. Ağın test sonucu performans parametreleri Tablo 8'de gösterilmiştir.

**Tablo 8. Burdur Fay Bölgesi YSA Test Sonuçları**

| Parametre | Değer | Yüzde |
|---|---|---|
| TP | 2 | |
| TN | 107 | |
| FP | 4 | |
| FN | 9 | |
| $P_0$ | 0,922414 | 92,24 |
| $P_1$ | 0,333333 | 33,33 |
| $S_n$ | 0,181818 | 18,18 |
| $S_p$ | 0,963964 | 96,40 |
| Ortalama | | 60,04 |

241

Uygulanan test sonucunda $P_0$ değeri yaklaşık olarak %93 çıkmaktadır. Bu değer istenilen aralıktadır. Eğitim sonucu test verisi ağa verildiğinde ağ; 2 depremin olacağını doğru tahmin etmiştir. 107 depremin gerçekleşmeyeceğini doğru tahmin etmiştir. Buna karşılık 9 depremi tahmin edememiş ve 4 depremi yanlış tahmin etmiştir.

Büyük ve Küçük Menderes Bölgesi 6 Ekim 2010 ile 18 Aralık 2013 tarihleri arasındaki magnitüd değeri 2.9 ve üzerinde olan 122 adet deprem verisi de, YSA'nın test edilmesi amacıyla kullanılmıştır. Ağın test sonucu performans parametreleri Tablo 9'da gösterilmiştir.

**Tablo 9. Büyük ve Küçük Menderes Bölgesi YSA Test Sonuçları**

| Parametre | Değer | Yüzde |
|---|---|---|
| TP | 19 | |
| TN | 53 | |
| FP | 32 | |
| FN | 18 | |
| $P_0$ | 0,746479 | 74,65 |
| $P_1$ | 0,372549 | 37,25 |
| $S_n$ | 0,513514 | 51,35 |
| $S_p$ | 0,623529 | 62,35 |
| Ortalama | | 56,40 |

Test sonucunda, $P_0$ değeri yaklaşık olarak %75 çıkmaktadır. Ağ 19 depremin olacağını ve 53 depremin gerçekleşmeyeceğini doğru tahmin etmiştir. Buna karşılık 18 depremi tahmin edememiş ve 32 depremi yanlış tahmin etmiştir.

Çalışma kapsamında incelenen Gediz ve Alaşehir Graben Bölgesinin 10 Mayıs 2010 ile 5 Aralık 2013 tarihleri arasındaki magnitüd değeri 2.8 ve üzerinde olan 122 adet deprem verisi, YSA'nın test edilmesi amacıyla kullanılmıştır. Ağın test sonucu performans parametreleri Tablo 10'da gösterilmiştir.

**Tablo 10. Gediz ve Alaşehir Graben Bölgesi YSA Test Sonuçları**

| Parametre | Değer | Yüzde |
|---|---|---|
| TP | 0 | |
| TN | 90 | |
| FP | 12 | |
| FN | 20 | |
| $P_0$ | 0,818182 | 81,82 |
| $P_1$ | 0 | 0,00 |
| $S_n$ | 0 | 0,00 |
| $S_p$ | 0,882353 | 88,24 |
| Ortalama | | 42,51 |

Test sonuçları incelendiğinde ağ çıkışlarının eşik değerden düşük olduğu için $P_1$ değeri sıfır bulunmuştur. Bu durumun düzeltilmesi amacıyla ağın eğitim veri setine yüksek değerli çıkışa sahip 20

242

adet vektör eğitim seti içerisinden seçilerek eğitim setine tekrar eklenmiş ve eğitim seti vektör sayısı 142 yapılmıştır. 142 giriş vektörlü eğitim sonucu Tablo 11'de gösterilmiştir.

**Tablo 11. Gediz ve Alaşehir Graben Bölgesi Vektör Eklemeli YSA Eğitim Sonuçları**

| Parametre | Değer | Yüzde |
|---|---|---|
| TP | 8 | |
| TN | 97 | |
| FP | 3 | |
| FN | 14 | |
| $P_0$ | 0,873874 | 87,39 |
| $P_1$ | 0,727273 | 72,73 |
| $S_n$ | 0,363636 | 36,36 |
| $S_p$ | 0,97 | 97,00 |
| Ortalama | | 73,37 |

Eğitim verisi ile ağın test edilmesi sonucu oluşan performans değerleri ise Tablo 12'de gösterilmiştir.

**Tablo 12. Gediz ve Alaşehir Graben Bölgesi Vektör Eklemeli YSA Test Sonuçları**

| Parametre | Değer | Yüzde |
|---|---|---|
| TP | 5 | |
| TN | 87 | |
| FP | 15 | |
| FN | 15 | |
| $P_0$ | 0,852941 | 85,29 |
| $P_1$ | 0,25 | 25,00 |
| $S_n$ | 0,25 | 25,00 |
| $S_p$ | 0,852941 | 85,29 |
| Ortalama | | 55,15 |

Vektör ekleme işleminden sonra gerçekleştirilen test sonucunda $P_0$ değeri yaklaşık olarak %85 çıkmaktadır. Eğitim sonucu test verisi ağa verildiğinde ağ; 5 depremin olacağını doğru tahmin etmiştir. 87 depremin gerçekleşmeyeceğini doğru tahmin etmiştir. Buna karşılık 15 depremi tahmin edememiş ve 15 depremi yanlış tahmin etmiştir.

**SONUÇ VE DEĞERLENDİRME**

YSA, ortaya çıktığı 1940'lı yıllardan günümüze kadar hala gelişmekte olan bir tahmin yöntemidir. Başlangıçta sadece doğrusal problemlerin çözümünde kullanılan YSA, ilerleyen zamanlarda doğrusal olmayan problemlerin çözümünde de etkin bir şekilde kullanılmaya başlamıştır. Canlı bir sinir sisteminin biyolojik yapısı temel alınarak geliştirilen modeller, canlı hafıza yapısını örnek alarak ve modelleyerek öğrenme yöntemlerini geliştirmiştir. Bilim adamlarının farklı yaklaşımları, yapılarda ağların ortaya çıkmasını sağlamış ve problemlere özgü ağ yapıları geliştirilmiştir. Bu gelişmelerin paralelinde, depremlerin tahmininde de YSA kullanılmaya başlamıştır.

Depremler en önemli doğal afetlerden biridir ve her yıl depremler nedeniyle binlerce insan hayatını kaybetmektedir. Günümüzde ve geçmişte depremlerin önceden tahmin edilebileceği düşüncesi bilim adamları tarafından tartışılmış, tahmin modelleri geliştirilmiştir. Tahmin yöntemleri arasında anormal hayvan davranışlarından, elektromanyetik dalgaların incelenmesine, radon gazı yoğunluğunun ölçülmesine gibi birçok tahmin parametreleri oluşturulmuştur. Bilim adamları depremleri etkileyen parametreleri farklı yaklaşımlarla tespit etmeye çalışmışlardır. Aynı zamanda bir parametredeki değişim ile deprem verileri arasında ilişki kurup inceledikleri parametrenin deprem ile olan ilişkisini ortaya koymaya çalışmışlardır.

Bu bağlamda çalışmada, Türkiye'de Gutenberg-Richter ilişkisine bağlı ve deprem tahminlerinde kullanılan b değerini temel alan bir ileri beslemeli geri yayılımlı yapay sinir ağı geliştirerek, ileri tarihli olası depremlerin tahmin edilebilmesi amaçlanmıştır. Çalışma kapsamında Gölhisar Çameli, Burdur Fay Zonu, Büyük Küçük Menderes, Gediz ve Alaşehir Graben olmak üzere toplam 4 farklı bölgede 2013 yılından önceki kesme magnitüd değerinden büyük 122 adet deprem verisi test amacıyla alınmıştır. Test verilerinden önceki 122 adet deprem verisi ise eğitim amacıyla alınmıştır. Eğitim ve test verileri katalog verileri üzerinde çeşitli işlemler yapılarak elde edilmiştir.

Çalışmanın eğitim ve test sonuçları incelendiğinde; Gölhisar Çameli Bölgesi YSA eğitim sonuçlarına göre $P_0$ değeri 0,893 bir değerle istenilen aralıkta çıkmıştır. Bu bölgede ağ, 2 depremin olacağını doğru tahmin etmiştir. 101 depremin gerçekleşmeyeceğini doğru tahmin etmiştir. Buna karşılık 12 depremi tahmin edememiş ve 7 depremi yanlış tahmin etmiştir. Burdur Fay Bölgesi YSA eğitim sonuçlarında $P_0$ değeri yaklaşık olarak %85 çıkmaktadır. Bu değer istenilen aralıktadır. Bu bölgede ağ 1 depremin olacağını doğru tahmin etmiştir. 100 depremin gerçekleşmeyeceğini doğru tahmin etmiştir. Buna karşılık 18 depremi tahmin edememiş ve 3 depremi yanlış tahmin etmiştir. Büyük Küçük Menderes Bölgesi YSA eğitim sonuçlarına göre, ağın eğitiminde $P_0$ değeri yaklaşık olarak %80 çıkmaktadır. Bu değer istenilen aralıktadır. Eğitim sonucu eğitim verisi ağa verildiğinde ağ 14 depremin olacağını doğru tahmin etmiştir. 79 depremin gerçekleşmeyeceğini doğru tahmin etmiştir. Buna karşılık 21 depremi tahmin edememiş ve 8 depremi yanlış tahmin etmiştir. Gediz ve Alaşehir Graben Bölgesi YSA eğitim sonuçlarına göre Ağın eğitiminde $P_0$ değeri yaklaşık olarak %84 çıkmaktadır. Bu değer istenilen aralıktadır. Eğitim sonucu eğitim verisi ağa verildiğinde ağ, 2 depremin olacağını doğru tahmin etmiştir. 98 depremin gerçekleşmeyeceğini doğru tahmin etmiştir. Buna karşılık 20 depremi tahmin edememiş ve 2 depremi yanlış tahmin etmiştir.

Bölgelerin test sonuçları incelendiğinde eğitim sonuçlarına paralel bulgular elde edilmiştir. Fakat Gediz ve Alaşehir Graben Bölgesi YSA test sonuçlarında ağın $P_1$ değeri sıfır bulunarak tahmin sonucu alınamamıştır. Bu durumu düzeltmek için girdi ve performans parametrelerinin hesaplanmasında Reyes ve diğerleri (2013) tarafından önerilen bir yöntem dikkate alınarak ağın eğitim veri setine yüksek değerli çıkışa sahip 20 adet vektör eğitim seti içerisinden seçilerek eğitim setine tekrar eklenmiş ve eğitim seti vektör sayısı 142 yapılmıştır. Bu uygulamadan sonra ki test tahmin

244

sonuçlarında ağın $P_0$ değeri yaklaşık %85'le istenilen aralıkta çıkmıştır. Bu doğrultuda ağ, 5 depremin olacağını doğru tahmin etmiştir. 87 depremin gerçekleşmeyeceğini doğru tahmin etmiştir. Buna karşılık 15 depremi tahmin edememiş ve 15 depremi yanlış tahmin etmiştir.

Genel olarak tahmin sonuçları değerlendirildiğinde ağ bütün bölgelerde gerçekleşmeyecek depremleri yüksek oranda tahmin etmiştir. Gerçekleşeceğini tahmin ettiği deprem sayıları da belli bir oranda gerçekleşirken, tahmin edemediği ve yanlış tahmin ettiği deprem tahminleri de sonuçlarda mevcuttur. YSA genel anlamda istenilen aralıklarda eldeki veri tabanında istenilen tahminleri belli bir başarı oranına kadar yapmasına rağmen çok yüksek oranlı gerçekleşecek deprem tahmini sunamamıştır. Verilerin yapısı incelendiğinde lineer olmayan bir veri topluluğu ile bu tahmin sonuçları istenilen düzeydedir. Gelecek çalışmalarda kullanılan giriş parametrelerine daha farklı deprem parametreleri eklenerek daha tutarlı tahminler elde edilebilir.(örneğin Radon gazı yoğunluğu gibi).

Panakkat ve Adeli tarafından yapılan çalışmada farklı ağ modelleri ve aynı girdi parametreleri kullanılarak deprem tahmini gerçekleştirilmiştir. Aynı girdi parametreleri ve veriler Levenberg Marquart, Radyal Tabanlı ve Yinelenen Sinir Ağı modellerinde denenmiştir. Yaptıkları çalışmada deprem tahmini konusunda en iyi sonucu Yinelenen Sinir Ağı modeli vermiştir. Panakkat ve Adeli tarafından kullanılan depremsellik parametreleri ile bu çalışmada kullanılan parametreler farklıdır. Bu açıdan, depremsellik parametrelerinin belirlenmesi ve belirlenen her bir parametrenin etkisinin ortaya konması önemlidir.

245

246